\documentclass[fleqn,10pt]{article}
\usepackage[utf8]{inputenc}
\usepackage[T1]{fontenc}

\usepackage{color}
\usepackage{amsmath,amssymb}
\usepackage{graphicx}
\usepackage{bm}
\usepackage{here}
\RequirePackage[utf8]{inputenc}
\RequirePackage[english]{babel}
\usepackage{xurl}
\usepackage{hyperref}

\RequirePackage{mathptmx}   

\usepackage{authblk}
\RequirePackage[left=2cm,%
                right=2cm,%
                top=2.25cm,%
                bottom=2.25cm,%
                headheight=12pt,%
                letterpaper]{geometry}%
                
\RequirePackage{fancyhdr}  
\RequirePackage{lastpage}  
\title{\vspace{-1.5cm}Identifying key differences between linear stochastic estimation and neural networks for fluid flow regressions}

\author[1]{Taichi Nakamura}
\author[2,1]{Kai Fukami}
\author[1,*]{Koji Fukagata}

\affil[1]{Department of Mechanical Engineering, Keio University}
\affil[2]{Department of Mechanical and Aerospace Engineering, University of California, Los Angeles}
\affil[*]{fukagata@mech.keio.ac.jp}

\date{}
\begin{document}
\maketitle
\vspace{-1.2cm}

\begin{abstract}
\noindent
Neural networks (NNs) and linear stochastic estimation (LSE) have widely been utilized as powerful tools for fluid-flow regressions.
We investigate fundamental differences between them considering two canonical fluid-flow problems: 1. the estimation of high-order proper orthogonal decomposition coefficients from low-order their counterparts for a flow around a two-dimensional cylinder, and 2. the state estimation from wall characteristics in a turbulent channel flow.
In the first problem, we compare the performance of LSE to that of a multi-layer perceptron (MLP).
With the channel flow example, we capitalize on a convolutional neural network (CNN) as a nonlinear model which can handle high-dimensional fluid flows.
For both cases, the nonlinear NNs outperform the linear methods thanks to nonlinear activation functions.
We also perform error-curve analyses regarding the estimation error and the response of weights inside models.
Our analysis visualizes the robustness against noisy perturbation on the error-curve domain while revealing the fundamental difference of the covered tools for fluid-flow regressions.
\end{abstract}

\flushbottom
%
%
\thispagestyle{empty}


\section{Introduction}

Recent boom of neural networks (NNs) has expanded into fluid mechanics~\cite{BNK2020,FFT2020}.
Although these have still been staying at the fundamental level, NNs have indeed shown their great potentials for data estimation~\cite{FFT2019,FFT2021,FukamiNMI}, control~\cite{BEF2019}, reduced-order modeling~\cite{MFRFT2020}, and turbulence modeling~\cite{DIX2019}.
Despite the recent enthusiastic trends on the use of NNs, we still have to rely largely on linear theories because their generalizability and transparency are, for the moment, superior to NN-based methods.
This indicates that we may be able to obtain some clues from {the relationships between NNs and linear theories} to develop more advanced NN-based techniques for fluid flow analyses.

Motivated above, of particular interest here is an analogy between NNs and linear methods.
Milano \& Koumoutsakos~\cite{Milano2002} used an autoencoder (AE) to perform low-dimensionalizaton for {the B}urgers equation and a turbulent channel flow.
They also reported that the operation inside the multi-layer perceptron (MLP) based AE with linear activation functions is equivalent to that of proper orthogonal decomposition (POD)~\cite{Lumely1967}.
More recently, Murata et al.~\cite{MFF2019} investigated the analogy between a convolutional neural network (CNN)-based AE and POD.
The strength of the nonlinear activation function used inside NNs was also demonstrated.
From the perspective on state estimation, Nair \& Goza~\cite{nair2020leveraging} have recently compared the MLP, Gappy POD, and linear stochastic estimation (LSE) for the POD coefficient estimation from local sensor measurements of laminar wake of a flat plate.
\begin{figure}
	\vspace{0mm}
	\begin{center}
		\includegraphics[width=0.95\textwidth]{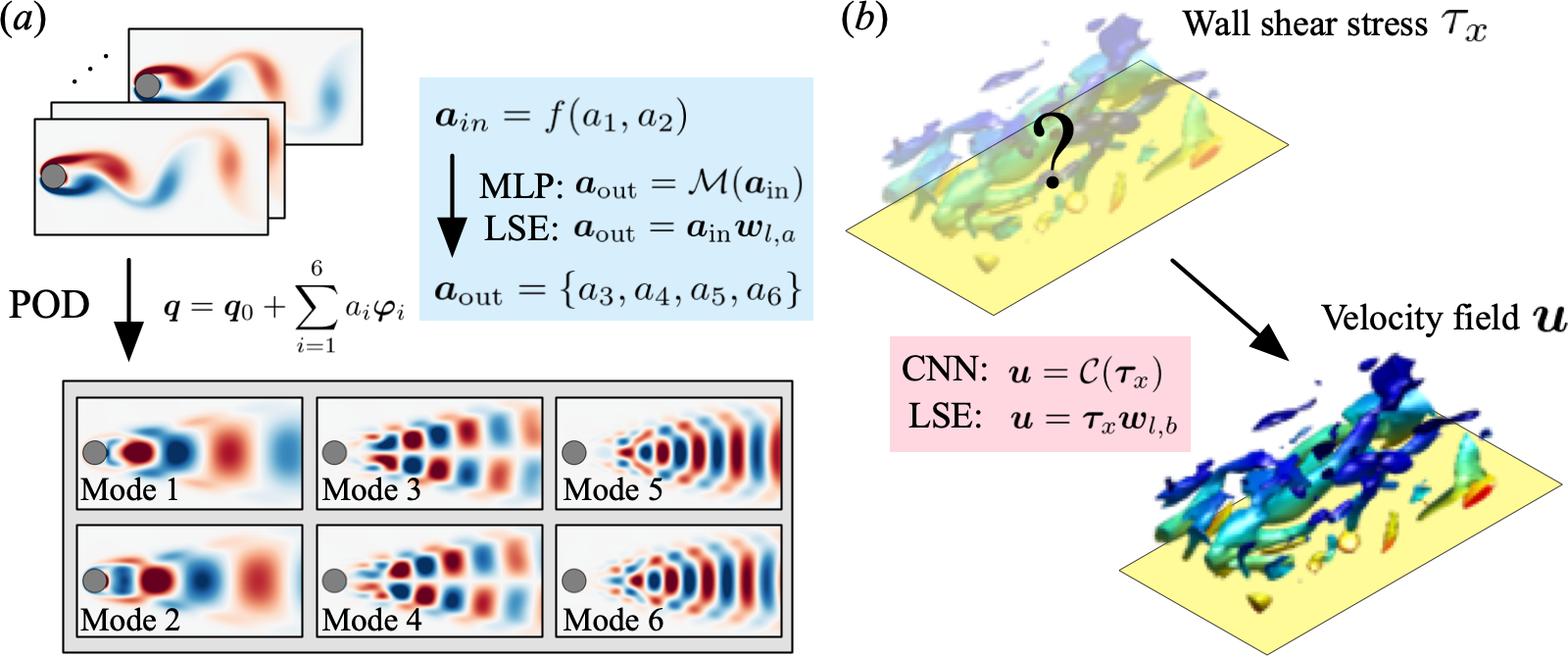}
		\caption{{Covered fluid flow regressions in the present study. $(a)$ POD coefficients estimation of a flow around a cylinder. $(b)$ State estimation in a turbulent channel flow.}}
		\label{fig_overivew}
	\end{center}
\end{figure}

To clarify the fundamental difference between linear methods and NNs, we here compare abilities of LSE, MLP, and CNN by considering two canonical fluid flow regression problems whose complexities are different with each other; 1. estimation of high-order POD coefficients from low-order counterparts of a flow around a cylinder~\cite{Loiseau2018}, and 2. state estimation from information on the wall in a turbulent channel flow~\cite{SH2017}, as illustrated in figure~\ref{fig_overivew}.
In particular, we seek key attributes for their estimations by focusing on influence{s} of biases inside NNs, optimization {methods,}
and robustness {against} noisy inputs.



\section{Regression methods}
\subsection{Multi-layer perceptron}

Let us first introduce a multi-layer perceptron (MLP)~\cite{RHW1986}.
The MLP, which mimics 
{the neurons in}
human's brain, has widely been utilized in physical sciences~\cite{Domingos2012}.
In the present paper, the MLP is used for low-dimensional fluid flow regressions, i.e., POD coefficient estimation, as an example of NN.
The MLP is 
{an}
aggregate of 
{the}
minimum unit{s called} 
perceptron{s}.
A linear superposition of input data $\bm q$ with weights $\bm{W}$ are passed inside a perceptron while 
{biases $\bm b$ being added,}
and then a nonlinear activation function $\phi$ is 
{applied}
such that
\begin{eqnarray}
    q_i^{(l)}=\phi\biggl(\sum_j W^{(l)}_{ij}q_j^{(l-1)}+b_j^{(l)}\biggr),
    \label{eq:mlp}
\end{eqnarray}
where $l$ denotes a layer index.
Weights among all edges $W_{ij}$ are optimized 
{with back{-}propagation~\cite{KB2014}}
{so as to minimize}
a loss function ${E}$.
The present MLP has hidden units of 4-8-16-8, while the number output nodes is 4 (3rd to 6th POD modes).
{T}he number of input nodes varies depending on 
{considered}
cases, 
{whose}
details will be provided in section \ref{Re_ex1}.
{In the simplest case,}
the present MLP ${\cal M}$ attempts to output ${\bm a}_{\rm out}=\{a_3,a_4,a_5,a_6\}$ from 
{2}
input{s} ${{\bm a}_{\rm in}=f(a_1,a_2)}$, 
{and}
the problem setting regarding weights ${\bm w}_m$ {(representing both ${\bm W}$ and ${\bm b}$)} inside the MLP can be represented as
\begin{eqnarray}
{\bm w}_m={\rm argmin}_{{\bm w}_m}||{\bm a}_{\rm out}-{\mathcal M}({\bm a}_{\rm in};{\bm w}_m)||_2.    
\end{eqnarray}
We use the $L_2$ 
{error} 
norm as a loss function.
Note that penalization terms, e.g., Lasso and Ridge penalties, are not considered in the present loss function because of the difficulty and cost in tuning the hyperparameter for regularization~\cite{NF2021}.


\subsection{Convolutional neural network}
\label{sec:cnn}

One of the issues associated with the MLP is that the number of edges inside it may explode {due to its fully-connected structure when handling}
high-dimensional data such as fluid flows. 
To overcome this issue in dealing with fluid flow problems, a convolutional neural network (CNN)~\cite{LBBH1998} has widely been accepted as a good candidate~\cite{FFT2019,MFZNF2021}.
We capitalize on the combination of two- and three-dimensional CNNs for the state estimation task in the present study.
The convolutional layer, which is a fundamental operation inside CNNs, extracts spatial features from input data using a filter operation,
\begin{eqnarray}
    q^{(l)}_{ijm} = {\phi}\biggl({b_m^{(l)}}+\sum^{K-1}_{k=0}\sum^{{H}-1}_{p=0}\sum^{{H}-1}_{s=0}h_{p{s}km}^{(l)} q_{i+p{{-C}}\,j+{s{-C}}\,k}^{(l-1)}\biggr),
    \label{eq:2DCNN}
\end{eqnarray}
where {$C={\rm floor}(H/2)$,} $K$ is the number of filters in a convolution layer, and {$b_m^{(l)}$ is the bias}.
This filter operation achieves efficient data handling for two- or three-dimensional flow fields~\cite{FFT2020}.
For the three-dimensional CNN, the three-dimensional convolution is performed similarly to equation~\ref{eq:2DCNN}.
In the present paper, the size $H$ is set to 5.
In the CNN, the output from the filtering operation is passed through an activation function $\phi$, analogous to the operation inside an MLP (equation~\ref{eq:mlp}).
Filters $h$ are optimized with back-propagation to minimize a loss function.

We use the combination of two- and three-dimensional CNNs to output a three-dimensional turbulent state ${\bm u}$ from streamwise wall-shear stress ${\bm \tau}_{x,\rm wall}$ (details will be provided in section \ref{Re_ex2}).
The optimization problem of weights ${\bm w}_c$ 
{(representing both ${\bm h}$ and ${\bm b}$)} 
for the CNN ${\cal C}$ can be expressed as
\begin{eqnarray}
{\bm w}_c={\rm argmin}_{{\bm w}_c}||{\bm u}-{\mathcal C}({\bm \tau}_{x,\rm wall};{\bm w}_c)||_2.    
\end{eqnarray}

\begin{table*} 
    \caption{{Structure of {2D-3D CNN} ${\cal C}$ for turbulent channel flow example.}}
    \begin{center}
    \begin{tabular}{cccc}
        \hline
        Layer & Data size  & Layer & Data size \\ \hline
        Input ${\bm \tau}_{x,\rm wall}$ & $(32,32,1)$ & 14th Conv. & $(5,5,16)$  \\
        1st 2D Conv. & $(32,32,8)$ & 15th 2D Conv. & $(32,32,32)$ \\
        2nd 2D Conv. & $(32,32,8)$ & 16th 2D Conv. & $(32,32,32)$\\
        3rd 2D Conv. & $(32,32,8)$ & Reshape & $(32,32,32,1)$\\
        4th 2D Conv. & $(32,32,8)$ &  1st 3D Conv. & $(32,32,32,8)$\\
        5th 2D Conv. & $(32,32,8)$ & 2nd 3D Conv.  & $(32,32,32,8)$\\
        6th 2D Conv. & $(32,32,16)$ & 3rd 3D Conv.  & $(32,32,32,16)$\\
        7th 2D Conv. & $(32,32,16)$ & 4th 3D Conv. & $(32,32,32,16)$ \\
        8th 2D Conv. & $(32,32,16)$ & 5th 3D Conv. & $(32,32,32,8)$\\
        9th 2D Conv. & $(32,32,16)$ & 6th 3D Conv. & $(32,32,32,8)$\\
        10th 2D Conv. & $(32,32,16)$ & 7th 3D Conv. & $(32,32,32,8)$\\
        11th 2D Conv. & $(32,32,16)$ & 8th 3D Conv. & $(32,32,32,8)$\\
        12th 2D Conv. & $(32,32,16)$ & 9th 3DConv. & $(32,32,32,3)$\\
        13th 2D Conv. & $(32,32,16)$ & Output ${\bm u}$ & $(32,32,32,3)$\\
        \hline
    \end{tabular}
    \end{center} \label{tab1}
    \vspace{-3mm}
\end{table*}

Again, the aim of this study is the comparison between LSE and NN. 
To achieve a fair comparison, it is ideal to consider 
{the}
same amount of weights for both LSE and NNs.
To this end, we perform singular value decomposition (SVD) for the LSE weights ${\bm w}_l$ to align the number of weights contained inside the CNNs and the LSE.
The details for the weight design for the LSE will be provided in the next section.
The operation can be expressed as ${\bm w}_l = {\bm U}{\bm \Gamma}{\bm V}^T$, where ${\bm U}\in {\mathbb{R}}^{n_{\rm input}\times n_{\rm rank}}$ and ${\bm V}^T\in {\mathbb{R}}^{n_{\rm rank}\times n_{\rm output}}$ are singular vectors, and ${\bm \Gamma}(\in {\mathbb{R}}^{n_{\rm rank}\times n_{\rm rank}})$ is a diagonal matrix whose diagonal elements are singular values.
Since the rank of LSE weights in the present study is $n_{\rm rank}={\rm rank}({\bm w}_l)=193$ according to our preliminary test, the present number of LSE weights based on SVD reduction $n_{{\bm w}_{\rm LSE, SVD}}$ is 197\,632 $(= n_{\rm input}\times n_{\rm rank} = 1024 \times 193)$.
Capitalizing on this SVD-based weight reduction, we are now able to obtain the target number of weights of approximately 197\,000 to determine the CNN parameters.
In this study, our CNN contains 196\,608 weights with the parameters in table~\ref{tab1}.

\subsection{Linear stochastic estimation}

For comparison to the NNs, we use linear stochastic estimation (LSE)~\cite{AM1988,SH2017}.
In this study, we express target data $\bm Q \in {\mathbb{R}}^{n_{\rm data}\times n_{\rm output}}$ (output) as a linear map ${\bm w}_l \in {\mathbb{R}}^{n_{\rm input}\times n_{\rm output}}$ with respect to input data $\bm P \in {\mathbb{R}}^{n_{\rm data}\times n_{\rm input}}$ such that ${\bm Q} = {\bm P}{\bm w}_l$, where $n_{\rm data}$ represents the number of training snapshots, $n_{\rm input}$ represents the number of input attribute, and $n_{\rm output}$ represents the number of output attribute.
Analogous to the optimization for NNs, the linear map ${\bm w}_l$ can be obtained through a minimization {such that}
\begin{equation}
    {\bm w}_l = {\rm argmin}_{{\bm w}_l}\parallel {\bm Q}-{{\bm w}_l}{\bm P}\parallel_2 {=({\bm P}^{\rm T}{\bm P})^{-1}{\bm P}^{\rm T}{\bm Q}}.
    \label{eqn_Linear}
\end{equation}
Note that penalization terms are not considered in the present loss function for the fair comparison to the NNs in terms of weight updating.
We also emphasize that the LSE is analytically optimized solving equation \ref{eqn_Linear} while the NNs are numerically optimized through back-propagation.
Hence, we can also compare the NNs and the LSE regarding the difference of the optimization approaches.
The optimized weights ${\bm w}_l$ can then be applied to test data.

\section{Results}

\subsection{Example 1: POD coefficient of two-dimensional cylinder wake at $Re_D=100$}
\label{Re_ex1}


As presented in figure~\ref{fig_overivew}$(a)$, we first aim to estimate high-order POD coefficients ${\bm a}_{\rm out}=\{a_3,a_4,a_5,a_6\}$ of a two-dimensional cylinder wake at $Re_D=100$ from information of low-order counterparts ${\bm a}_{\rm in}=f(a_1,a_2)$ such that ${\bm a}_{\rm out}={\mathcal F}_1({\bm a}_{\rm in})$, where ${\cal F}_1$ denotes a model for this purpose.
The LSE and the MLP are used as the model ${\cal F}_1$.
Flow snapshots are generated using a two-dimensional direct numerical simulation (DNS). 
The governing equations are the incompressible Navier--Stokes equations,
\begin{align}
    &\bm{\nabla} \cdot \bm{u}=0, ~~{\partial_t\bm{u}} + \bm{\nabla} \cdot (\bm{uu})  = - \bm{\nabla} p + {{Re}^{-1}_D}\nabla ^2 \bm{u},
\end{align}
where $\bm{u}$ and $p$ denote the velocity vector and the pressure, respectively.
All quantities are non-dimensionalized using the fluid density, the free-stream velocity, and the cylinder diameter.
The size of the computational domain is $(L_x, L_y)=(25.6, 20.0)$, and the cylinder center is located at $(x, y)=(9,0)$.
The grid spacing and the time step are respectively $\Delta x=\Delta y = 0.025$ and $\Delta t=2.5\times 10^{-3}$, while imposing the no-slip boundary condition on the cylinder surface using an immersed boundary method~\cite{kor2017}.
The number of grid points used for the DNS is $(N_x, N_y)=(1024, 800)$.
For the POD, the vorticity field $\omega$ around the cylinder is extracted as a domain of $8.2 \leq x \leq 17.8$ and $-2.4 \leq y \leq 2.4$ with $(N_x^*, N_y^*)=(384, 192)$.  
\begin{figure}[t]
	\vspace{0mm}
	\begin{center}
		\includegraphics[width=0.95\textwidth]{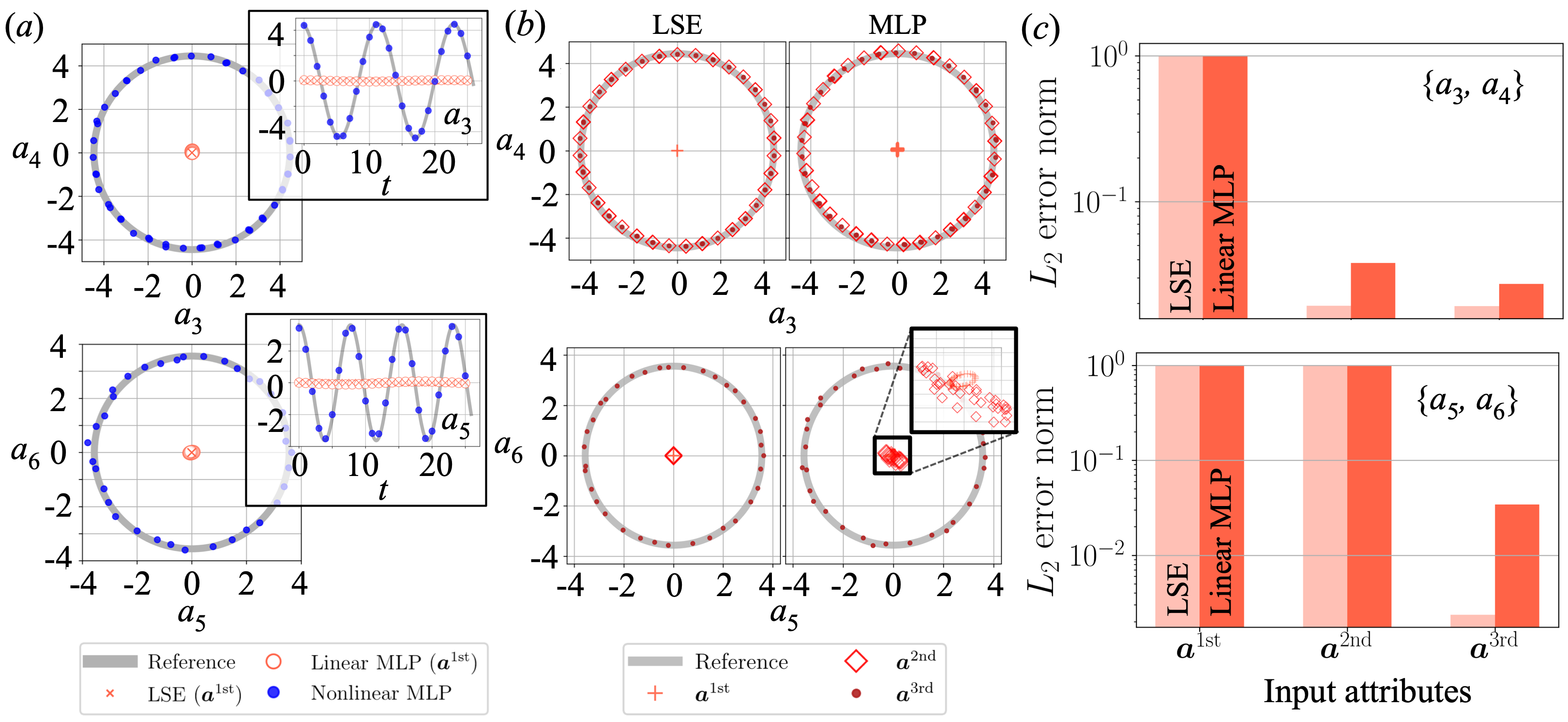}
		\caption{POD coefficients estimation from $(a)$ 1st order coefficients ${\bm a}_{\rm in}=\{a_1,a_2\}$ and $(b)$ 2nd \& 3rd order coefficients, and $(c)$ the $L_2$ error norm $\epsilon = ||{\bm a}_{\rm out, ref}-{\bm a}_{\rm out, est}||_2/||{\bm a}_{\rm out, ref}||_2$.
		}
		\label{fig4}
	\end{center}
\end{figure}
\begin{figure}[H]
	\begin{center}
		\centering\includegraphics[width=0.5\textwidth]{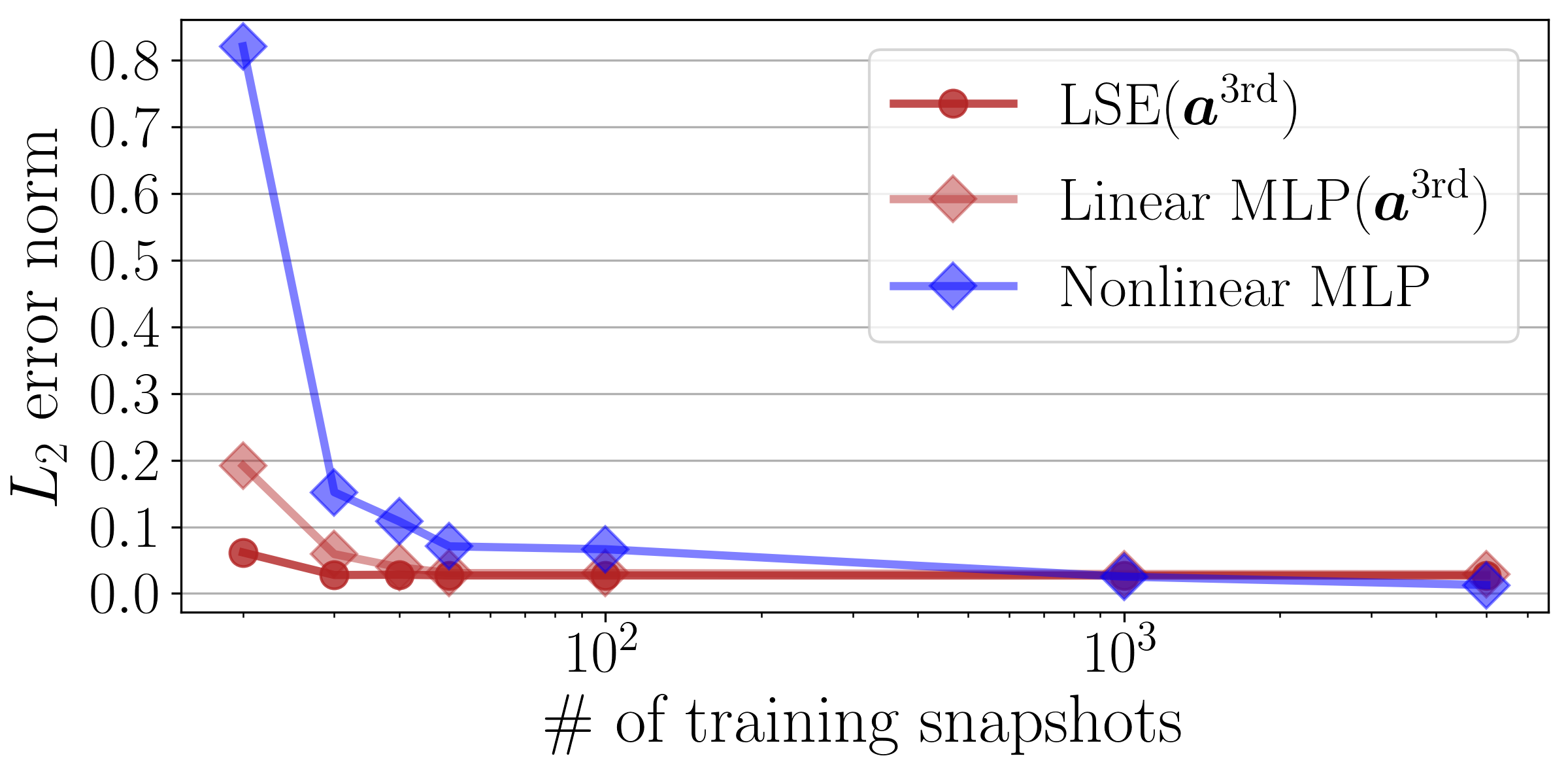}
		\caption{
		Dependence of the $L_2$ error norm for the POD coefficient estimation on the number of training snapshots.
		}
		\label{figA}
	\end{center}
\end{figure}
\begin{figure}[t]
	\begin{center}
		\includegraphics[width=0.90\textwidth]{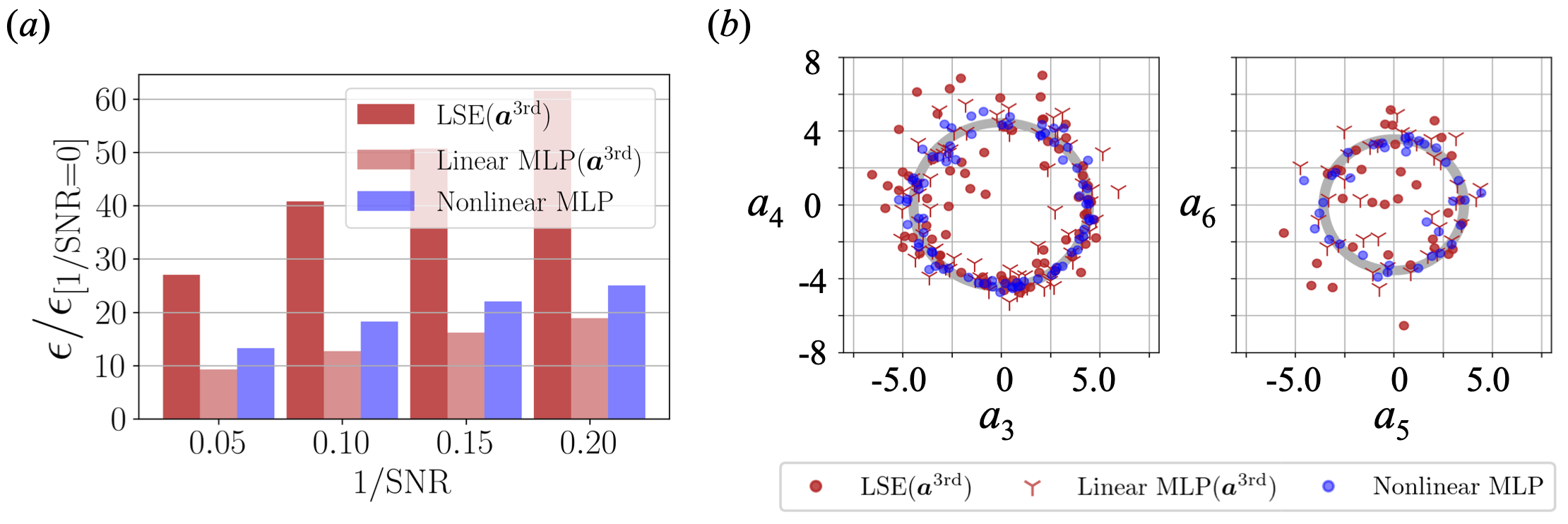}
		\caption{{
		Robustness for noisy input of POD coefficients estimation. $(a)$ Dependence of the increase ratio of the $L_2$ error norm $\epsilon/\epsilon_{\rm [1/SNR=0]}$ on noise magnitude. $(b)$ ${\bm a}_{\rm out}$ with ${\rm 1/SNR}=0.1$ for the linear models with ${\bm a}_{\rm in}={\bm a}^{\rm 3rd}$ and the nonlinear MLP with ${\bm a}_{\rm in}={\bm a}^{\rm 1st}$.}
		}
		\label{fig5}
	\end{center}
\end{figure}

We then take the POD for the collected snapshots to decompose the flow field $\bm q$ as ${\bm q}={\bm q}_0+\sum_{i=1}^{M}{a_i}{\bm \varphi}_i$, where $\bm \varphi$ denotes a POD basis, $a$ is the POD coefficient, ${\bm q}_0$ is the temporal average of the flow field, and $M$ represents the number of POD modes.
For training the present MLP and LSE, we use 5000 snapshots.
For comparison with LSE, we do not divide the training data for MLP into training and validation.
We also consider additional 5000 snapshots for the assessment.
We here compare the LSE 
{with}
the linear MLP and the nonlinear MLP with ReLU 
{activation}
function~\cite{NH2010}.
The ReLU is known as a good candidate to prevent a vanishing gradient issue.
We consider three patterns of input {${\bm a}_{\rm in}=f(a_1,a_2)$} for the LSE and the linear MLP; 
{the input $f(a_1,a_2)$} is ${\bm a}^{\rm 1st}=\{a_1,a_2\}$, or ${\bm a}^{\rm 2nd}=\{a_1,a_2,a_1a_2,a_1^2,a_2^2\}$, or ${\bm a}^{\rm 3rd}=\{a_1,a_2,a_1a_2,a_1^2,a_2^2,a_1^2a_2,a_1a_2^2,a_1^3,a_2^3\}$ while using only ${\bm a}^{\rm 1st}$ with the nonlinear MLP, 
Since Loiseau et al.~\cite{Loiseau2018} reported that the high-order coefficients ${\bm a}_{\rm out}$ can be represented using the quadratic expression of $a_1$ and $a_2$ due to its triadic interaction, this analysis enables us to check two viewpoints; 1. 
{whether the}
nonlinear function inside MLP works 
{to capture such nonlinear interactions,}
and 2. 
{whether the}
linear models can also be utilized 
{if}
a proper input 
{including the essential combinations of nonlinear terms is given.}

\begin{figure}[H]
	\begin{center}
		\includegraphics[width=0.9\textwidth]{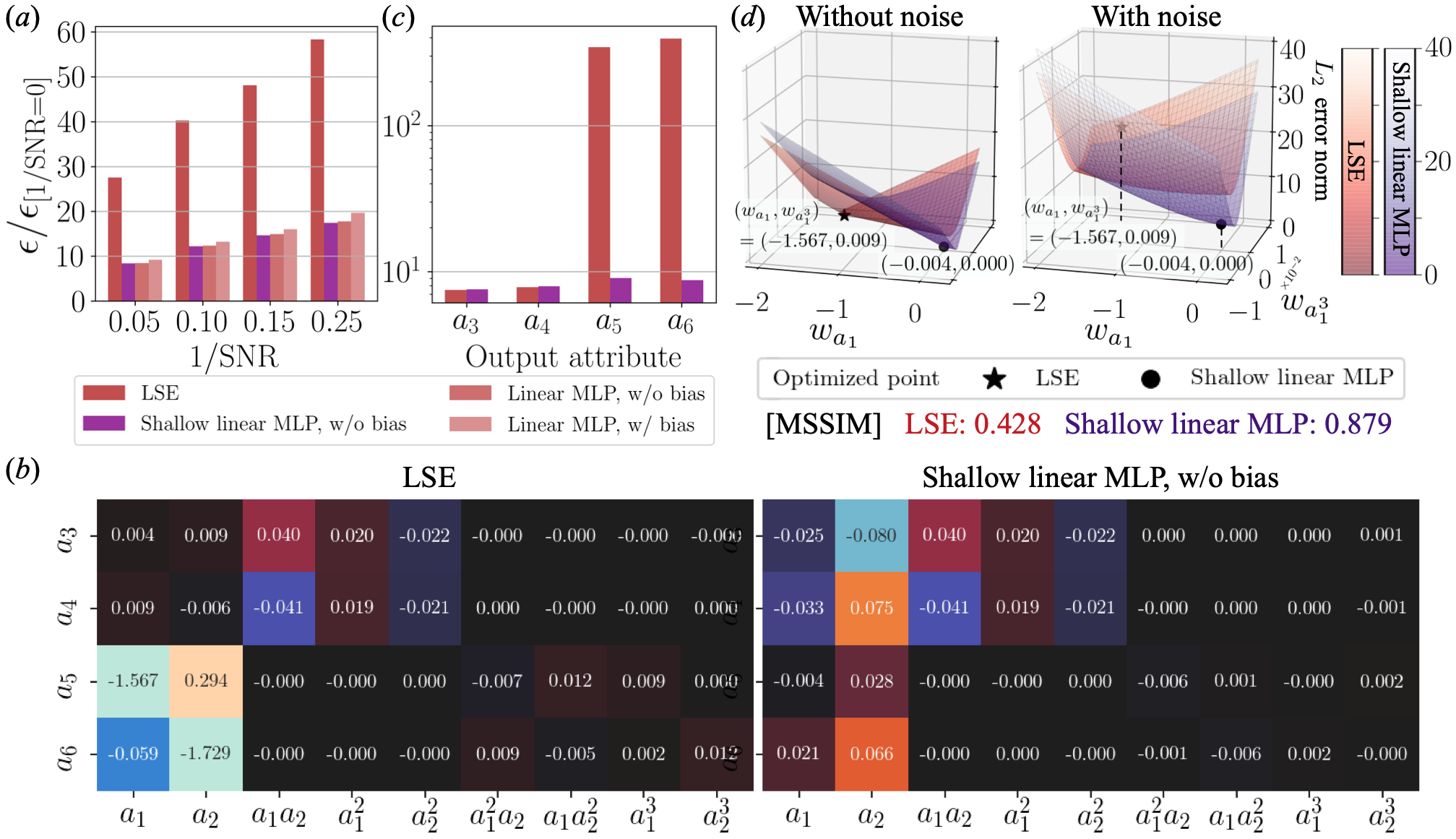}
		\caption{Comparison of the LSE and the linear MLP focusing on the bias and the optimization method. $(a)$ Increasing ratio $\epsilon/\epsilon_{\rm [1/SNR=0]}$ of the $L_2$ error norm $\epsilon$ from the original error $\epsilon_{\rm [1/SNR=0]}$ without the noisy input. $(b)$ Weight values of the LSE and the shallow linear MLP. $(c)$ Dependence of the increasing ratio $\epsilon/\epsilon_{\rm [1/SNR=0]}$ of the $L_2$ error norm at $1/{\rm SNR}=0.05$ on the output POD coefficients. $(d)$ Visualization of the error 
		{surface} 
		around the optimized point.}
		\label{fig5-1}
	\end{center}
\end{figure}
{
Let us demonstrate the estimation of ${\bm a}_{\rm out}=\{a_3,a_4,a_5,a_6\}$ from only the information of first-order coefficients ${\bm a}^{\rm 1st}=\{a_1,a_2\}$ such that ${\bm a}_{\rm out}={\cal F}_1({\bm a}^{\rm 1st})$, as shown in figure~\ref{fig4}$(a)$.
The nonlinear MLP shows its clear advantage against the LSE and the linear MLP for both coefficient maps.
The $L_2$ error norm $\epsilon = ||{\bm a}_{\rm out, ref}-{\bm a}_{\rm out, est}||_2/||{\bm a}_{\rm out, ref}||_2$ for each case are 1.00 (LSE), 1.00 (linear MLP), and 0.0119 (nonlinear MLP), respectively.}
This suggests that the nonlinear activation function plays an important role in estimation.
Noteworthy here, 
{however, is}
that this 
{nonlinearity}
can be 
{recovered}
by giving a proper input, i.e., ${\bm a}_{\rm in}=\{{\bm a}^{\rm 2nd},{\bm a}^{\rm 3rd}\}$, even if we only use the linear methods, as presented in figures~\ref{fig4}$(b)$ and $(c)$.
The reasonable estimation for $\{a_3,a_4\}$ can be achieved utilizing the input up to the 2nd order term ${\bm a}^{\rm 2nd}$, while that for $\{a_5,a_6\}$ requires the 3rd order term ${\bm a}^{\rm 3rd}$ with both the LSE and the linear MLP.
This trend is analogous to the observation by Loiseau et al.\cite{Loiseau2018}, as introduced above.
The LSE outperforms the linear MLP with the high-order coefficient inputs in this example, as shown in figure \ref{fig4}$(c)$; however, they will show a significant difference in terms of noise robustness as discussed later.

We then compare the LSE and the MLP in terms of the availability of training data.
The dependence of the $L_2$ error on the number of training snapshots is examined in figure~\ref{figA}.
Based on the results in figure~\ref{fig4}, we choose the third-order coefficients ${\bm a}_{\rm in} = {\bm a}^{\rm 3rd}$ for the linear models and the first-order coefficients for the nonlinear MLP, as the input for the models.
The LSE shows its advantage over the MLP models when the number of training snapshots is limited.
This is because the degree of freedom in the MLP is larger than that in the LSE.
Note that the MLP will show a clear advantage against the LSE in terms of noise robustness, which reveals the fundamental difference between the linear MLP and the LSE.

We here consider the Gaussian white noise defined by the signal-to-noise ratio (SNR), ${\rm SNR} = {\sigma^2_{\rm data}}/{\sigma^2_{\rm noise}}$,
where $\sigma{^2}_{\rm data}$ and $\sigma{^2}_{\rm noise}$ are 
{the} {variances} of input data and noise, respectively.
The behaviors for noisy inputs are summarized in figure \ref{fig5}.
As the linear models, we use the LSE and the linear MLP with ${\bm a}_{\rm in}={\bm a}^{\rm 3rd}$.
For comparison, we also monitor the nonlinear MLP with ${\bm a}_{\rm in}={\bm a}^{\rm 1st}$.
The response of LSE is much more sensitive than that of the covered MLPs.
This is caused by two considerable reasons: one is the influence of biases contained in the MLPs, as expressed in equation~\ref{eq:mlp}, while the other being the difference of optimization methods.
Hereafter, we will seek which has the main contribution for the noise robustness.


\begin{figure}
	\begin{center}
		\includegraphics[width=0.9\textwidth]{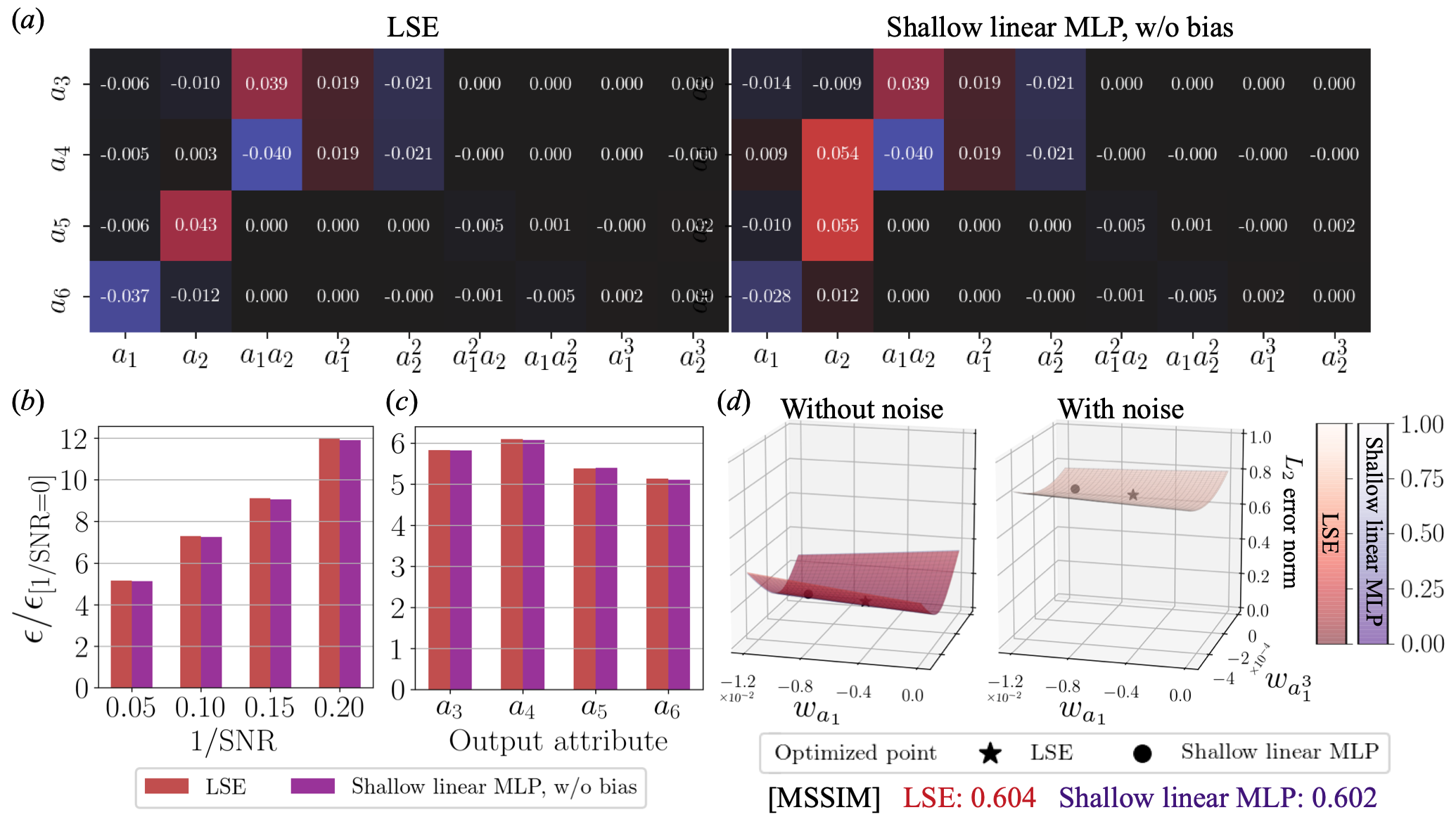}
		\caption{Comparison of the LSE and the linear MLP trained with the noisy training data. $(a)$ Weight values of the LSE and the shallow linear MLP. 
		$(b)$ Increasing ratio $\epsilon/\epsilon_{\rm [1/SNR=0]}$ of the $L_2$ error norm $\epsilon$ from the original error $\epsilon_{\rm [1/SNR=0]}$ without the noisy input. 
		$(c)$ Dependence of the increasing ratio $\epsilon/\epsilon_{\rm [1/SNR=0]}$ of the $L_2$ error norm at $1/{\rm SNR}=0.05$ on the output POD coefficients.
		$(d)$ Visualization of the error 
		{surface}
		around the optimized point.}
		\label{fig5-2}
	\end{center}
\end{figure}
For the investigation of the main contribution as mentioned above,
we consider the LSE and three types of MLPs as follows:
\begin{enumerate}
    \item LSE model: the same LSE model as that used above.
    \item Linear MLP model with bias ${\cal M}_1$: the same linear MLP as that used above
    \item Linear MLP model without bias ${\cal M}_2$: the biases are removed from the model ${\cal M}_1$ to investigate the influence on the bias.
    \item Shallow linear MLP model without bias ${\cal M}_3$: the MLP with a single hidden layer is prepared to align the number of weights with the LSE so that the difference of optimization methods can be assessed.
\end{enumerate}

The dependence of the $L_2$ error norm on the noise magnitude of each model is summarized in figure~\ref{fig5-1}$(a)$.
There is almost no difference among the covered MLP models.
This suggests that the bias and the number of layers do not contribute to the noise robustness so much.
The other observation here is that the shallow linear MLP is still more robust than the LSE, even though the model structures are identical with each other.
To examine this point, we visualize the weights inside the LSE and the shallow linear MLP in figure~\ref{fig5-1}$(b)$.
The weights for the second-order term input ($a_1a_2,a_1^2, a_2^2$) are optimized to the same values with each other, while that for the first-order input term input exhibits a significant difference.
This is caused by the difference of the optimization methods.
Moreover, this point is examined by visualizing an error surface for the input of $a_1$ and $a_1^3$ with the output of $a_5$, as shown in figure~\ref{fig5-1}$(d)$.
The reason for the choice of this input-output combination is that the output $a_5$ is one of the most sensitive components to the noise for the LSE, as presented in figure~\ref{fig5-1}$(c)$.
The optimized solutions are different to each other, which is likely caused by the difference in optimization methods.
What is notable here is that the noise addition drastically changes the 
{error-surface}
shape 
of the LSE, while that of the MLP 
{changes only slightly.}
This 
{difference}
can be 
{quantified}
using 
{the}
mean structural similarity index measure (MSSIM)~\cite{WBSS2004}.
We apply this to the elevation of each error 
{surface,}
i.e., without $E_r\in\mathbb{R}^{M\times N}$ and with noise 
$E_n\in\mathbb{R}^{M\times N}$, where $M,N$ are the number of samples on the error 
{surface}
for each weight axis.
MSSIM is mathematically expressed as 
\begin{eqnarray}
    {\rm MSSIM}(E_r,E_n)= \frac{1}{M^\prime N^\prime}\sum_{i=1}^{M^\prime}\sum_{j=1}^{N^\prime}{\rm SSIM}(e_{r,ij},e_{n,ij}),~~~~~~{\rm SSIM}(e_r, e_n) = \frac{(2\mu_r\mu_n+C_1)(2\sigma_{rn}+C_2)}{(\mu_r^2+\mu_n^2+C_1)(\sigma_r^2+\sigma_n^2+C_2)}.
    \label{eq:ssim}
\end{eqnarray}
This measurement can assess a similarity between two images $E_r\in\mathbb{R}^{M\times N}$ and $E_n\in\mathbb{R}^{M\times N}$ by considering their mean $\mu$ and standard deviation $\sigma$.
To obtain MSSIM, the SSIM
{in} 
a small window of two images $e_r\in\mathbb{R}^{m\times n}$ and $e_n\in\mathbb{R}^{m\times n}$, where $M^\prime=M-m+1$ and $N^\prime=N-n+1$, 
{is computed}
and its average is taken
{over the image}.
As the constant values $C_1$ and $C_2$ in equation~\ref{eq:ssim}, we set $\{C_1,C_2\}=\{0.16,1.44\}$ following Wang et al.~\cite{WBSS2004}.
As presented in figure~\ref{fig5-1}$(d)$, the MSSIM of the shallow linear MLP reports 0.879 while that of LSE is 0.428, which 
{indicates}
that 
{the deformation of}
the error 
{surface}
{is substantially larger in the LSE.}
Due to this large deformation of the error surface, 
{the optimum}
point 
{of the LSE}
is pushed up vertically
{in the error space of figure~\ref{fig5-1}$(d)$}.
This indicates that the weights obtained by the LSE in an analytical manner guarantee the global optimal solution over the training data; however, 
{this solution}
may not be 
optimal 
from the viewpoint of noise robustness.
On the other hand, the MLP provides a noise-robust solution, although it is not the {exact} global 
{optimum}
over the training data since the MLP weights are a numerical solution 
{obtained}
through back{-}propagation.

Considering the characteristics of LSE which fits training data, we then add noisy data of ${\rm SNR}=0.05$ to training data for both LSE and the shallow linear MLP (without bias), as summarized in figure \ref{fig5-2}.
As shown in figure~\ref{fig5-2}$(a)$, there is almost no difference in terms of weight values despite that there was in figure~\ref{fig5-1}.
The similar observations can also be found for several analyses in figures~\ref{fig5-2}$(b)-(d)$.
These suggest that the LSE can also obtain robustness by adding noise to the training data.
Note that the increasing ratio $\epsilon/\epsilon_{\rm [1/SNR=0]}=5$ in figure~\ref{fig5-2}$(b)$ is not quite large because the original error $\epsilon_{\rm [1/SNR=0]}$ is small 
($\epsilon_{\rm [1/SNR=0]}=4.49\times10^{-2}$ for the LSE and $\epsilon_{\rm [1/SNR=0]}=4.52\times10^{-2}$ for the shallow linear MLP).

\begin{figure}[H]
	\vspace{0mm}
	\begin{center}
		\includegraphics[width=0.95\textwidth]{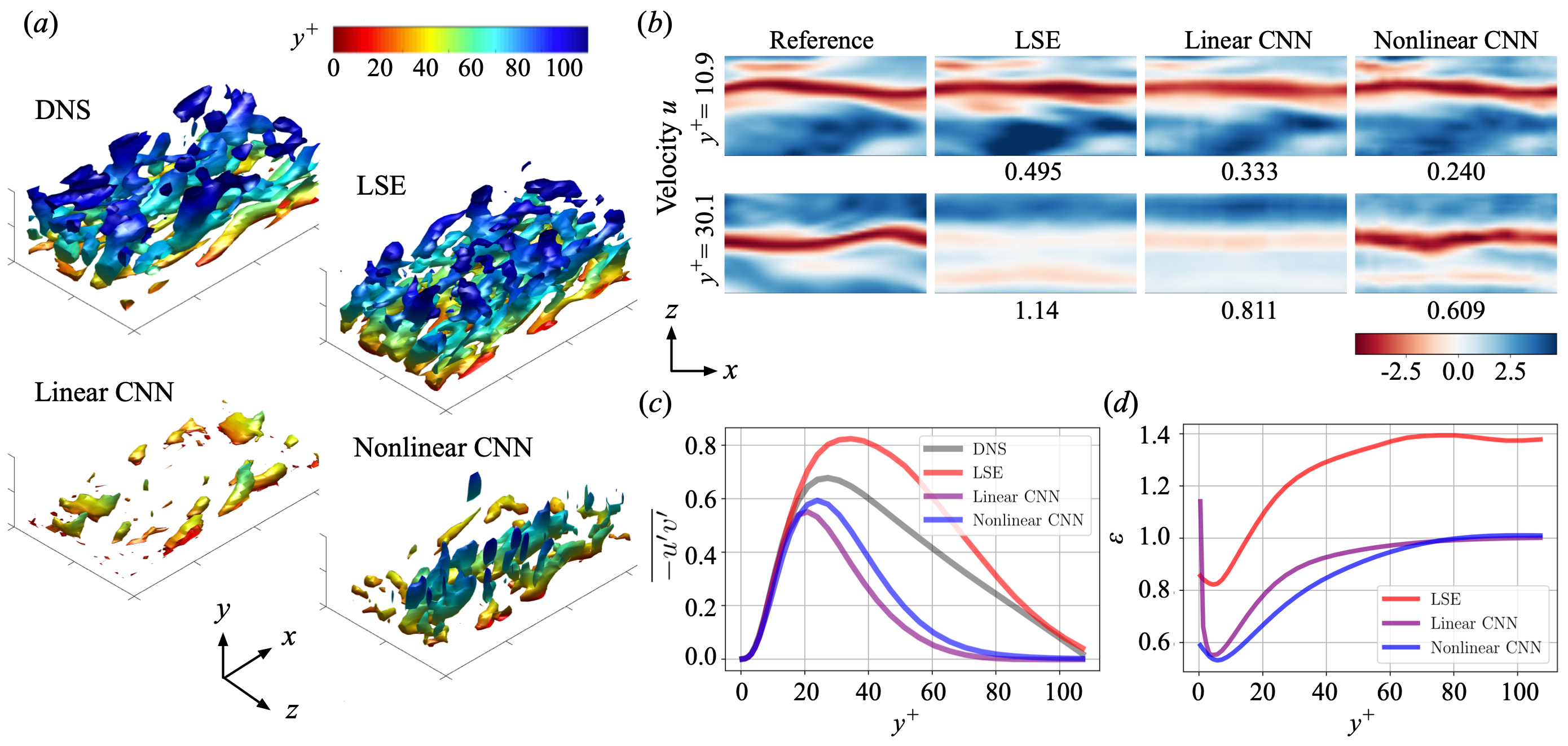}
		\caption{{
		Estimation of turbulent channel flow from streamwise wall-shear stress. 
		$(a)$ Isosurfaces of the $Q$ criterion $(Q^+ = -0.005)$.
		$(b)$ $x-z$ sectional velocities at $y^+=10.9$ and $30.1$. The values underneath the contours report the $L_2$ error norm for each velocity attribute. 
		$(c)$ Reynolds shear stress $\overline{-u^\prime v^\prime}$.
		$(d)$ Dependence of the ensemble $L_2$ error norm over three velocity components on the $y$ position.
		}}
		\label{fig6}
	\end{center}
\end{figure}

\subsection{Example 2: velocity field in a minimal turbulent channel flow at $Re_\tau=110$}
\label{Re_ex2}

{
We then perform the comparison of NNs and LSE for a more complex problem.
Let us consider the estimation of a velocity field ${\bm u}$ in a minimal turbulent channel flow at $Re_{\tau}=110$ from the streamwise wall shear stress input ${\bm \tau}_x$, as illustrated in figure \ref{fig_overivew}$(b)$, such that ${\bm u}=\mathcal{F}_2({\bm \tau}_x)$, where $\mathcal{F}_2$ denotes a model for example 2.
As already mentioned, we use LSE and CNN as the model $\mathcal{F}_2$.
The training data is generated using a three-dimensional DNS which numerically solves the incompressible Navier--Stokes equations,
\begin{equation}
    \bm{\nabla} \cdot {\bm u} = 0,~~~{ {\partial_t {\bm u}} + \bm{\nabla} \cdot ({\bm u \bm u}) =  -\bm{\nabla} p  + {{Re}^{-1}_\tau}\nabla^2 {\bm u}},
\end{equation}
where ${\bm u}$ and $p$ represents the velocity vector and pressure, respectively\cite{FKK2006,NFHNF2021}.
The quantities are non-dimensionalized with the channel half-width $\delta$ and the friction velocity $u_\tau$.
The computational domain is $(L_{x}, L_{y}, L_{z}) = (\pi\delta, 2\delta, 0.5\pi\delta)$ with the number of grid points of $(N_{x}, N_{y}, N_{z}) = (32, 64, 32)$.
The grid is arranged uniformly in the $x$ and the $z$ directions, while nonuniformly in the $y$ direction.
The time step is $\Delta t^+=0.0385$, where the subscript $+$ denotes the wall units.
We use 10\,000 snapshots to train the models.
For equivalent comparison with LSE, we do not divide the training data into training and validation.
We also prepare additional 2700 snapshots for the assessment.
}

{
The channel flow fields estimated by the models are assessed in figure \ref{fig6}$(a)$.
Note again that the number of weights inside the CNN and the LSE is almost 
{the}
same 
{with}
each other 
as 
{explained}
in section \ref{sec:cnn}.
We here also consider the linear CNN (i.e., the same CNN structure, but with the linear activation function) to examine whether the similar observation to the linear MLP of the cylinder example
can be found or not even for a CNN whose filter operation is different from a fully-connected MLP.
The estimated fields are visualized using the second invariant of the velocity gradient tensor 
$Q{^+}$ of $-0.005$.
The field reconstructed by the LSE shows the similar behavior to the reference DNS data qualitatively, e.g., the amount of vortical structure; however, it should be emphasized that the LSE field {merely} provides turbulent-like structure, which is 
{different from that of} the DNS.

To investigate this point, we also visualize the $x-z$ sectional streamwise velocity distributions in figure~\ref{fig6}$(b)$.
Notably, the nonlinear CNN outperforms the LSE and the linear CNN, which is not intuitive from the observation with the $Q$ isosurfaces.
Especially, the advantage of the nonlinear 
{method}
can be 
{clearly}
found at $y^+=30.1$.
%
%
We also present the 
{time-averaged}
Reynolds shear stress ${-}\overline{u^\prime v^\prime}$ in figure~\ref{fig6}$(c)$.
The LSE curve 
{looks to be}
in 
{reasonable}
agreement 
{in its shape}
{(}although overestimated{)}, despite its high-error level as stated above.
With the observation for the reasonable reconstruction of LSE in figure~\ref{fig6}$(a)$, it implies that the flow field estimated by the LSE model is similar to the DNS data in a time-ensemble sense, although it does not match for each instantaneous field.

We also investigate the dependence of estimation on the $y$ position in figure \ref{fig6}$(d)$.
Analogous to previous studies~\cite{CHBH2006,SH2017}, it is hard to estimate 
{the velocity field in}
the {region}
away from the wall because of the lack of correlation between the wall shear stress $\tau_x$ and the velocity field $\bm u$ away from the wall.
Within the range that can be estimated, the nonlinear CNN presents the better estimation than the linear methods in terms of both the $L_2$ error and the statistics.
}

\begin{figure}[t]
	\vspace{0mm}
	\begin{center}
		\includegraphics[width=1.00\textwidth]{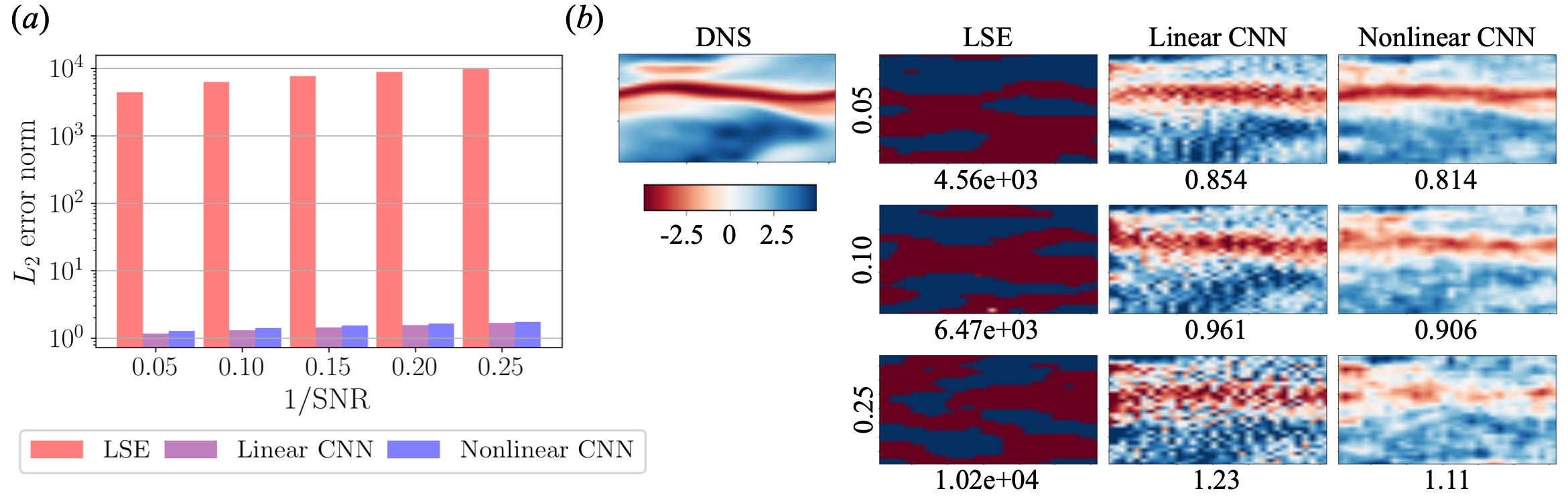}
		\caption{{
		Robustness for noisy input 
		{in the}
		turbulent channel flow example. $(a)$ $\epsilon/\epsilon_{\rm [1/SNR=0]}$ on noise magnitude. $(b)$ Contours of estimated streamwise velocity fluctuation $u^{\prime}$ at $y^+=10.9$. The values underneath the contours indicate the ensemble $L_2$ error norm over three velocity components. The numbers on the side of the contour represent $1/{\rm SNR}$.}}
		\label{fig7}
	\end{center}
\end{figure}
\begin{figure}
	\vspace{0mm}
	\begin{center}
		\includegraphics[width=1.0\textwidth]{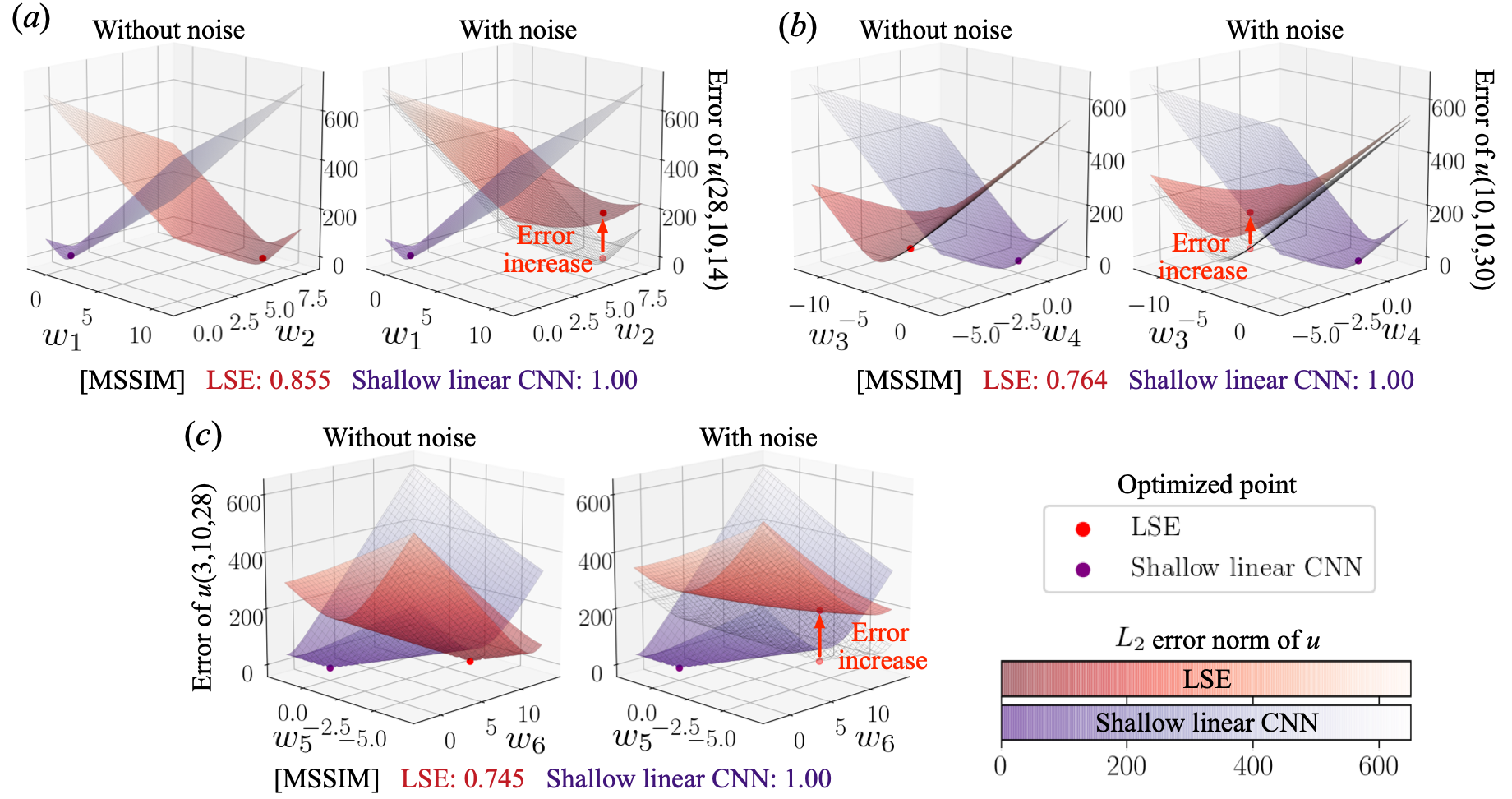}
		\caption{
		{Error-surface}
		analysis 
		{in the}
		turbulent channel flow example. The axis $x$ and $y$ are arbitrary weights picked up from each model.}
		\label{fig8}
	\end{center}
\end{figure}

We further assess the difference among the linear and nonlinear methods focusing on noise robustness, as summarized in figure \ref{fig7}.
Analogous to the noise investigation for the POD coefficient estimation, the noisy input for the the wall shear stress is designed with the SNR.
The $L_2$ error of the LSE explodes rapidly with the noise addition, which is the same behavior as that in figure~\ref{fig5}$(a)$.
In contrast, the CNNs are still able to reconstruct the large scale structure despite that the LSE cannot even with $1/{\rm SNR}=0.05$.
These findings also make us to suspect the role of optimization methods.

Let us then discuss the contribution of optimization methods for noise robustness.
We here skip the influence on bias uses since there is no significant effects according to our preliminary test, which is akin to the trend with the cylinder example.
For the particular demonstration here, we arrange a shallow linear CNN, in which there are $n_{\rm output}$ filters and each filter has a shape of $(n_{\rm input},1)$.
Note that this shallow CNN is distinct against the CNNs used above, which were based on the original SVD-based weight reduction expressed in section~\ref{sec:cnn}.
The use of the shallow linear CNN enables us to observe the weight sensitivity against noisy inputs while comparing to the LSE directly, thanks to its filter shape.

The error {surfaces} of the LSE and the shallow linear CNN are visualized in figures~\ref{fig8}$(a)$-$(c)$.
Note that we here choose six points for the visualization.
The error used in the error surface is arranged by the streamwise velocity $u$ at an arbitrary point in the $x-z$ cross section at $y^+=15.4$.
In the case of the shallow linear CNN, the noise has little influence on the error surface.
On the other hand, the error surfaces of the LSE drastically change their shape with the existence of noise.
These trends can also be found with the MSSIM.
It implies that the CNN can also obtain noise robustness thanks to the gradient method, while it is hard to obtain with the LSE, which is analogous to the observation with the POD coefficient estimation.

\section{Conclusions}
\label{sec:conclusion}

Fundamental differences between neural networks (NNs) and linear methods were investigated considering canonical fluid flow regression problems: 1. the estimation of POD coefficients of a flow around a cylinder, and 2. the state estimation from wall measurements in a turbulent channel flow.
We compared linear stochastic estimation (LSE) with multi-layer perceptron (MLP) and convolutional neural network (CNN).
For both regression problems, efficacy of nonlinear function can be observed.
We also found that a linear model could surrogate a nonlinear model by giving an appropriate combination of inputs under the consideration of nonlinear relationship.
This enables us to expect that the combination of nonlinear activation functions and proper inputs can further enhance the prediction capability of the model, which is similar to the observation in several previous studies~\cite{GH2017,KL2020,PC2021}.
In addition, the linear NNs were more robust against noise than the LSE, and the reason for this was revealed by visualizing the error surface.
The error surface told us that the difference in optimization methods has a significant contribution to the noise robustness.

Although we observed the strength of nonlinear NNs from several perspectives, we should note that the learning process of NNs can also be unstable depending on a problem setting since it is founded on a gradient method.
This implies that we may not reach a reasonable valley of a weight surface, especially when the problem has a multimodality and the availability of training data is limited~\cite{MFMVF2021}.
In this sense, the LSE can provide us a stable solution in a theoretical manner.
Hence, we may be able to unify these characteristics for further improvement of NN learning pipelines, e.g., transfer learning~\cite{MFZF2020,guastoni2020convolutional}, so that a learning process of NN can be started from a reasonable solution while achieving a proper noise robustness.
Furthermore, it should also be emphasized that we can introduce a loss function associated with {\it a priori} knowledge from physics since both NNs and LSE are based on a minimization manner in terms of weights.
Inserting a physical loss function may be one of the considerable paths towards practical applications of both methods~\cite{LY2019,raissi2017physics,DZ2021}.

\section*{Acknowledgement}

This work was supported through JSPS KAKENHI Grant Number 18H03758 and 21H05007 by Japan Society for the Promotion of Science.

\section*{Author contributions statement}

T.N and Ka.F designed research; T.N. performed research and analyzed data; Ka.F. and Ko.F. supervised. 
All authors reviewed the manuscript.

\end{document}